\documentstyle[12pt]{article}
\sf

\begin{document}
\input vatola.sty
\def\disp{\displaystyle}
\def\bm{\begin{minipage}[t]}
\def\em{\end{minipage}}
\def\spe#1{\special{BMF=#1.BMF}}
\def\fz{\footnotesize\zihao{6}}
\def\ref#1{\fz\par\hangindent\parindent\indent\llap{#1\enspace}\ignorespaces}
\def\srh{\stackrel{rightpoonup}}
\def\i{\int}
\def\s{\sum}
\def\f{\frac}
\def\p{\partial}
\def\bt{\begin{tabular}}
\def\et{\end{tabular}}
\def\md{\raisebox{-0.18\ccht}{\mbox{{\CC !{\char36}}}}}
\def\bfr{\begin{flushright}}
\def\mm{\mbox{\boldmath $ }}
\def\efr{\end{flushright}}
\def\bfl{\begin{flushleft}}
\def\efl{\end{flushleft}}
\def\vs{\vspace}
\def\hs{\hspace}
\def\sta{\stackrel}
\def\pb{\parbox}
\def\bc{\begin{center}}
\def\ec{\end{center}}
\def\sp{\setlength{\parindent}{2\ccwd}}
\def\bp{\begin{picture}}
\def\ep{\end{picture}}
\def\uni{\unitlength=1mm}
\def\REF#1{\par\hangindent\parindent\indent\llap{#1\enspace}\ignorespaces}

\noindent
\bc
{\LARGE\bf Different versions of perturbation expansion based on the
single-trajectory quadrature method}

\vs*{5mm}
{\large W. Q. Chao(Zhao)$^{1,~2}$ and C. S. Ju$^{2}$}

\vs*{11mm}
{\small \it 1. China Center of Advanced Science and Technology (CCAST)}

{\small \it (World Lab.), P.O. Box 8730, Beijing 100080, China}

{\small \it 2.  Institute of High Energy Physics, Chinese Academy of Sciences,

P. O. Box 918(4), Beijing 100039, China}
\ec
\vspace{2cm}
\begin{abstract}

The newly developed single trajectory quadrature method is applied
to a two-dimensional example. The results based on different versions of
new perturbation expansion and the new Green's function
deduced from this method are compared to each other,
also compared to the result
from the traditional perturbation theory. As the first application to
higher-dimensional non-separable potential the obtained result further
confirms the applicability and potential of this new method.

\end{abstract}
\vspace{1cm}

\noindent
PACS{:~~11.10.Ef,~~03.65.Ge}
\newpage

\section*{\bf 1. Introduction}
\setcounter{section}{1}
\setcounter{equation}{0}

Recently a new method has been developed in Refs.[1,2] to solve the
low-lying quantum wave functions of Schroedinger
equation using quadratures along a single trajectory. Based on the
expansion on $1/g$, where $g $ is a scale factor expressing the strength
of the potential, Schroedinger equation can be cast into
a series of first order partial differential equations, which is further
reduced to a series of integrable first order ordinary differential
equations by single-trajectory quadratures. New perturbation series
expansion and Green's functions of the wave equation are also derived
based on this method, both for one-dimensional and N-dimensional cases.
Some examples for one-dimensional problems have been illustrated in
[1,2]. Recently this new method has successfully applied to solve quantum dot
problem[3] and Yukawa potential[4].

In this paper, as the first complete illustration, this new method is applied to
higher dimensional problems. Specially, it is shown in the paper how to derive
the single trajectory based on Hamilton-Jacobi method. Schroedinger equation with
a two dimensional non-separable potential is solved using this
new method, based on two different versions of
perturbation series, exponential and polynomial expansion, and using the Green's
function deduced from this method.
The results show that these different versions are all equivalent. It is
also shown in the paper that the result of the new method is the same as
the one based on the traditional perturbation theory. It is much easier to
obtain higher order perturbed wave functions using this new method, compared to
the traditional perturbation theory where the calculation of the second
order perturbed wave function is already quite tedious.
These results further confirm the potential of this new method in the
future developments and applications.

The paper is organized in the following way. In Section II. A brief
review of the single trajectory quadrature method is given first. It is followed
by a complete illustration to solve Schroedinger equation with a simple
two-dimensional non-separable potential
$$
V=g^2[\f{1}{2}(x^2+b^2y^2)+\mu U(x,y)].
\eqno(1.1)
$$
The same problem is solved using exponential and polynomial
perturbation expansion, respectively, in Section III. In Section IV
the Green's function method derived based on the new method is applied to the
same example. Comparisons between different versions of this new method and to
the traditional perturbation theory are given in Section V, together with
a brief discussion and summary.

\newpage

\section*{\bf 2. Application to a two-dimensional example}
\setcounter{section}{2}
\setcounter{equation}{0}

The method newly developed in refs.[1,2]
provides a completely new way to solve the low-lying quantum wave
function for one particle N-dimensional Schroedinger equation
based on quadratures along a single trajectory.
$$
H\Phi ({\bf q})= E \Phi({\bf q}),
\eqno(2.1)
$$
where
$$
H=-\f{1}{2}\nabla^2+V({\bf q})
\eqno(2.2)
$$
is the Hamiltonian of a unit mass particle, and
$$
\nabla^2=\sum\limits^N_{i=1}\f{\partial^2}{\partial q_i^2}.
\eqno(2.3)
$$

The method consists three basic steps:\\

\noindent
1.  For a potential $ V({\bf q})\geq 0 $, a scale factor $ g^2 $ is
introduced as
$$
V({\bf q})=g^2v({\bf q}).
\eqno(2.4)
$$
Expressing $ \Phi({\bf q})=e^{-g{\bf S}({\bf q})} $, Both $g{\bf S}({\bf q})$
and the energy $ E $ are expanded in terms of $1/g$ in the following way.
$$
g{\bf S}({\bf q})=g{\bf S}_0({\bf q})+{\bf S}_1({\bf q})+
\f{1}{g}{\bf S}_2({\bf q})+\dots
\eqno(2.5)
$$
$$
E=gE_0+E_1+\f{1}{g}E_2+\cdots.
\eqno(2.6)
$$
Substitute the expansion (2.5) and (2.6) into the Schroedinger
equation (2.1) and equating the coefficients of $ g^{-n} $. After this
step the second order partial differential Schroedinger equation
changes to a series of first order partial differential equations of
$ \left\{{\bf S}_i\right\} $
and $ \left\{E_i\right\} $.

\noindent
2.  The lowest order equation of $ {\bf S}_0 $ in the series of
equations, i. e.,
$$
{(\nabla {\bf S}_0)}^2=2v
\eqno(2.7)
$$
is equivalent to a Hamilton-Jacobi equation in classical
mechanics with $-v$ as its potential and $ e=0^+ $ as the total energy. It
could be proved that the solution ${\bf S}_0$ can be written as
$$
{\bf S}_0({\bf q})=\int\limits^T_{T_0}[\f{1}{2}{\bf \dot{q}}^2-(-v({\bf q}))]dt,
\eqno(2.8)
$$
where the integral is along the trajectory $ {\bf q}(t) $ satisfying the
classical equations of motion
$$
{\bf \ddot{q}}(t)=\nabla v
\eqno(2.9)
$$
and the energy conservation
$$
\f{1}{2}{\bf \dot{q}}(t)^2-v({\bf q}(t))=0^+.
\eqno(2.10)
$$
The solution of (2.8) determines a single classical trajectory
${\bf S}_0$.\\

\noindent
3.  Introduce a new set of variables ${\bf S}_0$ and
$$
\alpha=(\alpha_1({\bf q}), \alpha_2({\bf q})\cdots\alpha_{N-1}({\bf q}))
\eqno(2.11)
$$
satisfying
$$
\nabla\alpha_j\cdot\nabla {\bf S}_0=0, ~~~~~~j=1,2,\cdots, N-1.
\eqno(2.12)
$$
Changing $ (q_1\dots q_N)\rightarrow ({\bf S}_0, \alpha_1,\dots,
\alpha_{N-1}) $, all $ \left\{{\bf S}_i(\bf q)\right\} $ become
functions of $ {\bf S}_0 $ and $ \alpha$, i. e, $\left\{{\bf S}_i
({\bf S}_0, \alpha)\right\} $. Based on (2.12) the series of first
order partial differential equations of $ \left\{{\bf S}_i\right\} $ reduces
to a series of first order ordinary differential equations with
variable $ {\bf S}_0 $. These equations can be solved by quadratures
along the single trajectory of constant $\alpha$, i. e, along
$ {\bf S}_0 $.

Now we turn to the example of two dimensional non-separable potential.
To solve Schroedinger equation
$$
H\Phi(x,y)=E\Phi(x,y)
\eqno(2.13)
$$
with
$$
\begin{array}{rcl}
H&=&-\f{1}{2}\nabla^2+ V(x,~y)\\
&&\nonumber\\
\nabla^2&=&{\displaystyle \f{\partial^2}{\partial x^2}}+
{\displaystyle \f{\partial^2}{\partial y^2}},
\end{array}
\eqno(2.14)
$$
where $V(x,~y)= g^2 v(x,~y)$ and
$$
v(x,~y)=\f{1}{2}(x^2+b^2y^2)+\mu U(x,~y).
\eqno(2.15)
$$
we follow the three steps of the new method.

\noindent
1.  Express $ \Phi=e^{-g{\bf S}} $ and introduce the expansion (2.5) and
(2.6) for $ g {\bf S} $ and $ E $.

Substituting (2.5) and (2.6) into (2.13) the series of equations for
 $ \left\{{\bf S}_i\right\} $ and $ \left\{E_i\right\} $ are obtained:
$$
\begin{array}{rl}
{(\nabla {\bf S}_0)}^2=&2v, \\
\nabla {\bf S}_0\cdot\nabla {\bf S}_1=&\f{1}{2}\nabla^2{\bf S}_0-E_0, \\
\nabla {\bf S}_0\cdot\nabla {\bf S}_2=&\f{1}{2}[\nabla^2{\bf S}_1-
{(\nabla {\bf S}_1)}^2]-E_1, \\
\nabla {\bf S}_0\cdot\nabla {\bf S}_3=&\f{1}{2}[\nabla^2{\bf S}_2-2
(\nabla {\bf S}_1)\cdot(\nabla {\bf S}_2)]-E_2,\\
&\vdots
\end{array}
\eqno(2.16)
$$

\noindent
2. $ {\bf S}_0 $ can be expressed as
$$
{\bf S}_0(x,y)=\int\limits^T_{T_0}[\f{1}{2}(\dot{x}^2+\dot{y}^2)-(-v(x,y))]dt
\eqno(2.17)
$$
and $ (x,y) $ satisfy the equations of motion:
$$
\left\{\begin{array}{rcl}
\ddot{x}&=&\f{\partial v}{\partial x}=x+\mu\f{\partial U}{\partial x}  \\
\ddot{y}&=&\f{\partial v}{\partial y}=b^2y+\mu\f{\partial U}{\partial y}.
\end{array}
\right.
\eqno(2.18)
$$

In the following we take $ U(x,y)=x^2y^2 $ as our example. To
solve (2.18), further expansion of $ x $ and $ y $ on $ \mu $ is
introduced:
$$
\left\{\begin{array}{rl}
x=&x_0+\mu x_1+\mu^2x_2+\cdots\\
y=&y_0+\mu y_1+\mu^2y_2+\cdots.
\end{array}\right.
\eqno(2.19)
$$
A series equations for different orders of $ \mu $ is obtained:
$$
\hs*{43mm}
\begin{array}{rll}
\mu^0:& \ddot{x}_0=x_0& \hs*{40mm} (2.20) \\
&\ddot{y}_0=b^2y_0& \\
\mu^1:&\ddot{x}_1=x_1+2 x_0y^2_0& \hs*{40mm} (2.21) \\
&\ddot{y}_1=b^2y_1+2 x^2_0y_0& \\
\mu^2:&\ddot{x}_2=x_2+2 x_1y^2_0+4x_0y_0y_1& \\
&\ddot{y}_2=b^2y_2+2 x^2_0y_1+4x_0x_1y_0& \hs*{40mm} (2.22)  \\
&\vdots& \\
\end{array}
$$

We first solve eq. (2.20) and obtain
$$
\left\{\begin{array}{rl}
x_0=&c_xe^t+d_xe^{-t} \\
y_0=&c_ye^{bt}+d_ye^{-bt}.\\
\end{array}
\right.
\eqno(2.23)
$$

Introducing the initial condition:
$$
\left\{\begin{array}{lr}
x=0& \\
& ~~~~~~~~~~~~~~~~~{\rm at}~~t\rightarrow -\infty \\
y=0& \\
\end{array}
\right.
\eqno(2.24)
$$
we have
$$
\left\{\begin{array}{rl}
x_0=&c_xe^t \\
y_0=&c_ye^{bt}. \\
\end{array}
\right.
\eqno(2.25)
$$
The undetermined constants $ c_x $ and $ c_y $ will be fixed
after solving the series of equations of $ \{x_i\} $ and
 $ \{y_i\} $,  by
fixing the end point of the trajectory $ (x_T, y_T) $ at final time $ T $.

Now we solve equation (2.21) for $ (x_1, y_1) $. Substituting (2.25)
into (2.21) we could obtain
$$
\left\{\begin{array}{rl}
x_1=&c_xc_y^2\f{1}{2b(b+1)}e^{(2b+1)t} \\
y_1=&c^2_xc_y\f{1}{2(b+1)}e^{(b+2)t}.
\end{array}
\right.
\eqno(2.26)
$$
Similarly, substituting (2.25) and (2.26) into (2.22) the solution could
be obtained as
$$
\left\{\begin{array}{rl}
x_2=&\f{c_xc^2_y}{b+1}[\f{c_x^2}{2(b+2)(b+1)}e^{(3+2b)t}+
\f{c_y^2}{8b^2(2b+1)}e^{(1+4b)t}] \\
y_2=&\f{c_yc^2_x}{b+1}[\f{c_x^2}{8(b+2)}e^{(4+b)t}+\f{c^2_y}{2b(b+1)(2b+1)}
e^{(2+3b)t}].
\end{array}
\right.
\eqno(2.27)
$$
In principle, the whole series of $ \left\{x_i\right\},\left\{y_i\right\} $
could be obtained by solving the series of equations in different
orders of $ \mu $. In the following we will study the result up to the order
of $ \mu^2, $ i.e.,
$$
\begin{array}{rl}
x=&x_0+\mu x_1+\mu^2 x_2  \\
y=&y_0+\mu y_1+\mu^2 y_2. \\
\end{array}
\eqno(2.28)
$$
Now using the condition that $ (x,y)=(x_T, y_T) $ at $ t=T $, we can fix
$ c_x $ and $ c_y $ up to $ \mu^2 $:
$$
\left\{\begin{array}{rl}
c_x=&x_Te^{-T}\left\{1-\mu\f{y^2_T}{2b(1+b)}+\mu^2\f{y^2_T}{b{(1+b)}^2}
[\f{x^2_T}{2+b}+\f{y^2_T(3b+1)}{8b(2b+1)}]\right\} \\
c_y=&y_Te^{-bT}\left\{1-\mu\f{x^2_T}{2(1+b)}+\mu^2\f{x^2_T}{{(1+b)}^2}
[\f{y^2_T}{2b+1}+\f{x^2_T(3+b)}{8(2+b)}]\right\}. \\
\end{array}
\right.
\eqno(2.29)
$$
It can be readily shown that the solution (2.28) satisfies the energy
conservation condition, namely
$$
\f{1}{2}(\dot{x}^2+\dot{y}^2)-[\f{1}{2}(x^2+b^2y^2)+\mu x^2y^2]=0^+.
\eqno(2.30)
$$
Substituting (2.25)--(2.29) into (2.17) and taking $ T_0=-\infty $ we
obtain
$$
{\bf S}_0=\f{1}{2}(x^2+by^2)+\f{\mu}{2(b+1)}x^2y^2-\f{\mu^2}{4{(b+1)}^2}
x^2y^2(\f{x^2}{b+2}+\f{y^2}{2b+1}).
\eqno(2.31)
$$

\noindent
3.  Now we try to solve $ \left\{{\bf S}_i\right\} $ and $ \left\{E_i
\right\} $ using the single trajectory quadrature along $ {\bf S}_0 $.
After $ {\bf S}_0 $ is obtained, making the variable transformation, the
series of equations (2.15) becomes a series of quadratures along the single
trajectory $ {\bf S}_0 $:
$$
\begin{array}{rl}
E_0=&\f{1}{2}\nabla^2{\bf S}_0|_{at~ q=0},\\
{\bf S}_1({\bf q})={\bf S}_1({\bf S}_0,\alpha)=&\int\limits^{{\bf S}_0}_0
\f{d{\bf S}_0}{{(\nabla {\bf S}_0)}^2}[\f{1}{2}\nabla^2{\bf S}_0-E_0], \\
E_1=&\f{1}{2}{[\nabla^2{\bf S}_1-{(\nabla{\bf S}_1)}^2]}_{at~ q=0}, \\
{\bf S}_2({\bf q})={\bf S}_2({\bf S}_0,\alpha)=&\int\limits^{{\bf S}_0}_0
\f{d{\bf S}_0}{{(\nabla {\bf S}_0)}^2}\{\f{1}{2}[\nabla^2{\bf S}_1-
{(\nabla{\bf S}_1)}^2]-E_1\}, \\
E_2=&\f{1}{2}{[\nabla^2{\bf S}_2-2(\nabla{\bf S}_1)\cdot(\nabla{\bf
S}_2)]}_{at~ q=0}. \\
&\cdots\\
\end{array}
\eqno(2.32)
$$
To perform the integration we first make the transformation
$$
\int\limits^{{\bf S}_0}_0\f{d{\bf S}_0}{{(\nabla {\bf S}_0)}^2}=
\int\limits^T_{-\infty}dt.
\eqno(2.33)
$$
This can be derived by directly substituting (2.31) into the left hand
side of (2.33).
By direct integration over t, we obtain the following result of
$ \{{\bf S}_i\} $ and $ \{E_i\} $ to the order of $ \mu^2 $ and $1/g$~:
$$
\begin{array}{rl}
E_0=&\f{1}{2}(1+b) \\
{\bf S}_1=&\f{\mu}{4(1+b)}(x^2+\f{y^2}{b})-\f{\mu^2}{4{(1+b)}^2}
(\f{x^4}{4(2+b)}+\f{x^2y^2}{b}+\f{9x^2y^2}{(2+b)(1+2b)}+
\f{y^4}{4b(1+2b)}) \\
E_1=&\f{\mu}{4b} \\
{\bf S}_2=&-\f{\mu^2}{16{(b+1)}^2}(x^2+\f{y^2}{b^3})-\f{\mu^2}{8b{(b+1)}^2}
(x^2+\f{y^2}{b}) \\
&-\f{\mu^2}{8{(1+b)}^2}\left\{\f{9}{(1+2b)(2+b)}(x^2+\f{y^2}{b})+\f{3}{2}
(\f{x^2}{2+b}+\f{y^2}{b^2(1+2b)})\right\} \\
E_2=&-\f{\mu^2}{16b^3(1+b)}(b^2+4b+1).
\end{array}
\eqno(2.34)
$$

Putting the above results together we have the approximate wave function and
energy of the ground state, up to the order of $ \mu^2 $ and $1/g$, as
follows:
$$
\begin{array}{rl}
\Phi(x,y)=&e^{-g{\bf S}_0-{\bf S}_1-g^{-1}{\bf S}_2} \\
E=&gE_0+E_1+g^{-1}E_2. \\
\end{array}
\eqno(2.35)
$$
The result shows clearly the two expansion series of $ \mu $ and $1/g$. In
principle, this solution could reach higher accuracy by treating higher
orders of $ \mu $ and $1/g$. This example gives us a general illustration of
how to solve the ground state wave function for high-dimensional
Schroedinger equation with non-separable potential.

\section*{\bf 3. Two versions of the new perturbation expansion series}
\setcounter{section}{3}
\setcounter{equation}{0}

For the same potential
$$
V(x,~y)=g^2[\f{1}{2}(x^2+b^2y^2)+\mu U(x,y)]
\eqno(3.1)
$$
the problem can also be solved by a new perturbation expansion based on
the new method. For this purpose, let us define
$$
g^2\mu U(x,y)=\epsilon U(x,y),
\eqno(3.2)
$$
i.e., $ \epsilon=g^2\mu $. Write
$$
V=V_0+\epsilon U
\eqno(3.3)
$$
and
$$
V_0=\f{1}{2}g^2(x^2+b^2y^2)=g^2v_0
\eqno(3.4)
$$
with
$$
v_0=\f{1}{2}(x^2+b^2y^2).
\eqno(3.5)
$$
The unperturbed Hamiltonian is
$$
H_0=-\f{1}{2}\nabla^2+V_0.
\eqno(3.6)
$$
It is very easy to solve
$$
H_0e^{-g{\bf S}_0}=E_0e^{-g{\bf S}_0}.
\eqno(3.7)
$$
The obtained trajectory is the same as (2.25), namely
$$
\left\{\begin{array}{rl}
&x=c_xe^t \\
&y=c_ye^{bt}, \\
\end{array}
\right.
\eqno(3.8)
$$
which satisfies the initial condition $ (x,y)=(0,0) $ at $ t=-\infty $.
The constants $ c_x $ and $ c_y $ are determined by the boundary
condition $ (x,y)=(x_T,y_T) $ at $ t=T $ and given as
$$
\left\{
\begin{array}{rl}
&c_x=x_Te^{-T} \\
&c_y=y_Te^{-bT}. \\
\end{array}
\right.
\eqno(3.8')
$$
(3.7) gives
$$
{\bf S}_0=\f{1}{2}(x^2+by^2), ~~~~E_0=\f{1}{2}(1+b).
\eqno(3.9)
$$
For
$$
H \Psi=(H_0+\epsilon U)\Psi=E\Psi
\eqno(3.10)
$$
we introduce
$$
\Psi=e^{-g{\bf S}_0-{\bf S}_1}\chi,
\eqno(3.11)
$$
then expand $ E $ and $ \chi $ on $1/g$. For the energy $ E $, we have the
same expansion as (2.6). There are two ways to expand $ \chi $, which will
be shown in the following.

\noindent
1. Exponential expansion of $ \chi $.

Introducing
$$ \chi=e^{-g^{-1}{\bf S}_2-g^{-2}{\bf S}_3-\cdots}.
\eqno(3.12)
$$
Substituting (3.11)-(3.12) into (2.10), a series of equations of $ \{{\bf S}_i\} $ and
$ \{E_i\} $ are obtained as follows:
$$
\begin{array}{rl}
{(\nabla{\bf S}_0)}^2=&2v_0=x^2+b^2y^2 \\
\nabla{\bf S}_0\cdot\nabla{\bf S}_1=&\f{1}{2}\nabla^2{\bf S}_0-E_0 \\
\nabla{\bf S}_0\cdot\nabla{\bf S}_2=&\f{1}{2}[\nabla^2{\bf S}_1-
{(\nabla{\bf S}_1)}^2]-E_1+\epsilon U \\
\nabla{\bf S}_0\cdot\nabla{\bf S}_3=&\f{1}{2}[\nabla^2{\bf S}_2-2
(\nabla{\bf S}_1)\cdot(\nabla{\bf S}_2)]-E_2\\
&\vdots
\end{array}
\eqno(3.13)
$$

This series of equations looks very similar to (2.15). However, there
are two differences between them. First, the $ {\bf S}_0 $ here is the
solution of (3.7), which is exact and very simple. Second, the equation
for $ {\bf S}_2 $ in the order of $g^0$ has a factor $ \epsilon U $ due
to the perturbation, while in (2.15) $ g^2\mu U$ is 1included in the
potential $ v $ in the equation of $ {\bf S}_0 $ in the order of $ g^2 $.
Now, we use the simple solution of $ H_0 $, i.e., (3.9), to define the
single trajectory $ {\bf S}_0=\f{1}{2}(x^2+by^2) $ and perform the same
procedure for (3.13), as we have done in Section II. To reach the same
accuracy we have to take the expansion to the orders of $ \epsilon^2 $
and $ g^{-5} $. In the calculation $ x $ and $ y $ in the integrand are
expressed as (3.8) and the integration $ \int^{{\bf S}_0}_0\f{d{\bf S}_0}
{{(\nabla{\bf S}_0)}^2} $ always transforms to $ \int^T_{-\infty}dt $.
For the simple example with $ U=x^2y^2 $, the obtained results are
$$
\begin{array}{rl}
{\bf S}_0=&\f{1}{2}(x^2+by^2) \hs*{20mm} E_0=\f{1}{2}(1+b) \\
{\bf S}_1=&0 \\
{\bf S}_2=&\f{1}{2(1+b)}\epsilon x^2y^2 \\
{\bf S}_3=&\f{\epsilon}{4(1+b)}(x^2+\f{y^2}{b}) \hs*{20mm}
E_3=\f{\epsilon}{4b} \\
{\bf S}_4=&-\f{\epsilon^2}{4{(1+b)}^2}x^2y^2(\f{x^2}{2+b}+\f{y^2}{1+2b}) \\
{\bf S}_5=&-\f{\epsilon^2}{4{(1+b)}^2}(\f{y^4}{4b(1+2b)}+\f{9x^2y^2}{(2+b)(1+2b)}
+\f{x^2y^2}{b}+\f{x^4}{4(2+b)}) \\
{\bf S}_6=&\f{-\epsilon^2}{8{(1+b)}^2}[\f{9}{(1+2b)(2+b)}(x^2+\f{y^2}b)
+\f{3}{2}(\f{x^2}{2+b}+\f{y^2}{b^2(1+2b)})] \\
&-\f{\epsilon^2}{8b{(1+b)}^2}(x^2+\f{y^2}{b})-\f{\epsilon^2}{16(1+b)^2}
(x^2+\f{y^2}{b^3})\\
E_6=&-\f{\epsilon^2}{16b^3(1+b)}(b^2+4b+1)
\end{array}
\eqno(3.14)
$$
Remembering $ \epsilon=g^2\mu $ the above results are the same as (2.34).

If we introduce
$$ g\lambda U=g^2\mu U
\eqno(3.15)
$$
instead of $ \epsilon U $, the same procedure could be taken. Only
the second and third equations in (3.13) should be changed, due to the
change of the order in $ g $ in the perturbation potential:
$$
\begin{array}{rcl}
\nabla{\bf S}_0\cdot\nabla{\bf S}_1&=&\f{1}{2}\nabla^2{\bf S}_0-E_0
+\lambda U \\
\nabla{\bf S}_0\cdot\nabla{\bf S}_2&=&\f{1}{2}[\nabla^2{\bf S}_1-(\nabla
{\bf S}_1)^2]-E_1
\end{array}
\eqno(3.16)
$$
The perturbation potential $ \lambda U $ now enters the equation of
$ {\bf S}_1 $ in $ g^1 $ order. To reach the same accuracy we
need here the expansion only to the order of $ \lambda^2 $ and $ g^{-3} $.
For $ U(x,y)=x^2y^2 $, the results are exactly the same as (3.14)
up to the order of $ \lambda^2 $ and $ g^{-3} $(i.e., $ \epsilon^2 $ and
$ g^{-5} $).

\noindent
2. Polynomial expansion.

Now, we introduce the expansion
$$
\chi=1+g{-1}\chi_1+g^{-2}\chi_2+\cdots
\eqno(3.17)
$$
and perform the same calculation for the perturbations $ \epsilon U $
and $ g\lambda U $, respectively. Substituting (2.6), (3.11) and
(3.17) into (3.10), comparing the coefficients of the same order of
 $ g^{-n} $, the series of equations of $ {\bf S}_0,{\bf S}_1, \{\chi_i\} $
and $ \{E_i\} $ could be obtained.

For the perturbation $ \epsilon U $ we have
$$
\begin{array}{rcl}
{(\nabla{\bf S}_0)}^2&=&2v_0 \\
\nabla{\bf S}_0\cdot\nabla{\bf S}_1&=&\f{1}{2}\nabla^2{\bf S}_0-E_0 \\
\nabla{\bf S}_0\cdot\nabla \chi_1 + \epsilon U &=& E_1 \\
\nabla{\bf S}_0\cdot\nabla \chi_2+\nabla{\bf S}_1\cdot\nabla \chi_1-\f{1}{2}
\nabla^2 \chi_1+\epsilon U \chi_1&=&E_1\chi_1+E_2 \\
\nabla{\bf S}_0\cdot\nabla \chi_3+\nabla{\bf S}_1\cdot\nabla \chi_2-\f{1}{2}
\nabla^2 \chi_2+\epsilon U \chi_2 &=& E_1\chi_2+E_2\chi_1+E_3 \\
\end{array}
\eqno(3.18)
$$
Based on the first equation of $ {\bf S}_0 $, the same trajectory of
$ {\bf S}_0=\f{1}{2}(x^2+by^2) $ is obtained. All $ \{\chi_i\} $ could be
solved by quadratures along the single trajectory $ {\bf S}_0 $, which is
again transformed to the integration of $ \int^T_{-\infty} dt $. For
$ U = x^2y^2 $, the final results up to the order of $ \epsilon^2 $
and $ g^{-5} $ are
$$
\begin{array}{rl}
{\bf S}_0=&\f{1}{2}(x^2+by^2), \hs*{30mm} E_0=\f{1}{2}(1+b) \\
{\bf S}_1=&0, \hs*{35mm} E_1=0  \\
\chi_1=&-\f{\epsilon}{2(1+b)}x^2y^2 \\
\chi_2=&-\f{\epsilon}{4(1+b)}(x^2+\f{y^2}{b})+\f{\epsilon^2}{8(1+b)} x^4y^4 \\
\chi_3=&\f{\epsilon^2}{8{(1+b)}^2}x^2y^2[\f{4+b}{2+b}x^2+\f{4b+1}{b(2b+1)}
y^2],~~~~~~~~ E_3=\f{\epsilon}{4b} \\
\chi_4=&\f{\epsilon^2}{1b{(1+b)}^2}\left\{\f{4+b}{2(2+b)}x^4+
\f{36}{(1+2b)(2+b)}x^2y^2+\f{5}{b}x^2y^2+
\f{4b+1}{2b^2(2b+1)}y^4\right\} \\
\chi_5=&\f{\epsilon^2}{16{(1+b)}^2}(x^2+\f{y^2}{b^3})+
\f{\epsilon^2}{8b{(1+b)}^2}
(x^2+\f{y^2}{b}) \\
&+\f{\epsilon^2}{8{(1+b)}^2}\left\{\f{9}{(1+2b)(2+b)}(x^2+\f{y^2}{b})+
\f{3}{2}(\f{x^2}{2+b}+\f{y^2}{b^2(1+2b)})\right\} \\
E_6=&-\f{\epsilon^2}{16b^3(1+b)}(b^2+4b+1) \\
\end{array}
\eqno(3.19)
$$
It is easy to prove that the polynomial expansion is equivalent to the
exponential one. When we expand (3.12) as $ \chi=1-g^{-1}{\bf S}_2-
g^{-2}{\bf S}_3\dots+\f{1}{2}{(-g^{-1}{\bf S}_2-g^{-2}{\bf S}_3-\cdots)}^2
+\cdots $ and compare to (3.17), we have
$$
\begin{array}{rcl}
\chi_1&=&-{\bf S}_2 \\
\chi_2&=&-{\bf S}_3+\f{1}{2}{\bf S}_2^2 \\
\chi_3&=&-{\bf S}_4+\f{1}{2}({\bf S}_2{\bf S}_3+{\bf S}_3{\bf S}_2)-
{\displaystyle \f{1}{3!}}{\bf S}^3_2 \\
&&\vdots \\
\chi_N&=&\sum\limits^N_{k=1}{(-1)}^k\f{1}{k!}\cdot
( \sum\limits_{(\sum\limits^k_{l=1}i_l)=N+k}
{\bf S}_{i_1}\cdot {\bf S}_{i_2}\dots {\bf S}_{i_k})
\end{array}
\eqno(3.20)
$$
Substituting (3.20) into (3.18) we return to (3.13) for $ {{\bf S}_2,
{\bf S}_3\cdots} $. In fact, when we expand $ e^{-g^{-1}{\bf S}_2
-g^{-2}{\bf S}_3\dots} $ according to (3.14), the obtained
expressions are equivalent to (3.19) up to the order of $ \epsilon^2 $ and
$ g^{-5} $.

Similarly, taking the perturbation $ g\lambda U $, the expansion gives
the following series of equations:
$$
\begin{array}{rcl}
{(\nabla{\bf S}_0)}^2&=&2v_0 \\
\nabla{\bf S}_0\cdot\nabla{\bf S}_1&=&\f{1}{2}\nabla^2{\bf S}_0-E_0+\lambda U\\
\nabla{\bf S}_0\cdot\nabla \chi_1
+\f{1}{2}[\nabla^2{\bf S}_1 -(\nabla{\bf S}_1)^2]
&=& E_1 \\
\nabla{\bf S}_0\cdot\nabla \chi_2+\nabla{\bf S}_1\cdot\nabla \chi_1-\f{1}{2}
\nabla^2 \chi_1
+\f{1}{2}[\nabla^2{\bf S}_1 -(\nabla{\bf S}_1)^2]
\chi_1
&=&E_1\chi_1+E_2 \\
\nabla{\bf S}_0\cdot\nabla \chi_3+\nabla{\bf S}_1\cdot\nabla \chi_2-\f{1}{2}
\nabla^2 \chi_2
+\f{1}{2}[\nabla^2{\bf S}_1 -(\nabla{\bf S}_1)^2]
\chi_2
&=& E_1\chi_2+E_2\chi_1+E_3 \\
&&\cdots\\
\end{array}
\eqno(3.21)
$$
Performing the same procedure for our
example with $ U=x^2y^2 $, expanding to the order of $ \lambda^2 $
and $ g^{-3} $, the obtained final results are equivalent to (3.19) for
$ g\lambda =\epsilon $.

\section*{\bf 4. The new Green's function method }
\setcounter{section}{4}
\setcounter{equation}{0}

Based on the new method the Green's functions are introduced in ref[2]
and applied to some one-dimensional examples. Now we are going to use
the new Green's function to solve our two dimensional
example. Assume that we have a two dimensional unperturbed Hamiltonian
$$
H_0=-\f{1}{2}\nabla^2+V_0(x,y)
\eqno(4.1)
$$
with $ V_0(x,y)=g^2v_0(x,y) $. The corresponding Schroedinger
equation
$$
H_0e^{-g{\bf S}_0}=gE_0e^{-g{\bf S}_0}
\eqno(4.2)
$$
can be solved and gives the unperturbed ground state wave function
 $ e^{-g{\bf S}_0} $ and energy $ gE_0 $. Introducing a perturbed
potential $ \epsilon U $, the Schroedinger equation
is now
$$
(H_0+\epsilon U)\Psi=(gE_0+\epsilon\Delta)\Psi
\eqno(4.3)
$$
The perturbed wave function $ \Psi=e^{-g{\bf S}_0} \chi $ can be obtained
using the following Green's function expression:[2]
$$
\Psi=e^{-g{\bf S}_0}+G\epsilon(-U+\Delta)\Psi,
\eqno(4.4)
$$
where
$$
G=e^{-g{\bf S}_0}C{(1+TC)}^{-1}e^{g{\bf S}_0}
\eqno(4.5)
$$
or
$$
\chi=1+C{(1+TC)}^{-1}\epsilon(-U+\Delta)\chi
\eqno(4.5')
$$
and $ T=-\f{1}{2}\nabla^2 $. To obtain the operator $ C $ in $ G $ we
have to start from the trajectory $ {\bf S}_0 $ which is obtained by
solving the unperturbed Schroedinger equation (4.2).
After the variable transformation $ (x,y)\rightarrow
({\bf S}_0,\alpha) $ under the condition $ \nabla{\bf S}_0\cdot\nabla
\alpha=0 $, the operator $ C $ is defined as
$$
C\equiv g^{-1}\theta~ h^2_{{\bf S}_0}
\eqno(4.6)
$$
where
$$
({\bf S}_0|\theta|{\bf S}^\prime_0)=\theta({\bf S}_0-{\bf S}^\prime_0)
=\left\{\begin{array}{lr}
1&~~~~~~{\bf S}_0>{\bf S}^\prime_0 \\
0&~~~~~~{\bf S}_0<{\bf S}^\prime_0
\end{array}\right.
\eqno(4.7)
$$
and
$$
h^2_{{\bf S}_0}={[{(\nabla{\bf S}_0)}^2]}^{-1}.
\eqno(4.8)
$$
When $ {\bf S}_0 $ is known the wave function $\Psi$ can be obtained
according to (4.4) and (4.5) by quadratures along the single trajectory
$ {\bf S}_0 $.

Now we turn to our example with
$$
v_0(x,y)=\f{1}{2}(x^2+b^2y^2).
\eqno(4.9)
$$
As shown in Section III, the trajectory is given by (3.8)-(3.9) as
$$
{\bf S}_0=\f{1}{2}(x^2+by^2)
\eqno(4.10)
$$
and
$$
\left\{\begin{array}{l}
x=x_Te^{-T}e^t \\
y=y_Te^{-bT}e^{bt}. \\
\end{array}\right.
\eqno(4.11)
$$
Substituting the above trajectory into (4.6)-(4.8), we have
$$
h^2_{{\bf S}_0}=\f{1}{x^2+by^2}
\eqno(4.12)
$$
and
$$
C=\f{1}{g}\int\limits^{{\bf S}_0}_0\f{d{\bf S}_0}{x^2+by^2}.
\eqno(4.13)
$$
Introducing the expressions of $ x(t) $ and $ y(t) $ in (4.11) into
(4.13), we finally have
$$
C=\f{1}{g}\int\limits^t_{-\infty}dt^\prime.
\eqno(4.14)
$$

Considering the simple example of $ \epsilon U=\epsilon x^2y^2 $, we
expand $ \chi $ as
$$
\chi=1+\sum\limits^\infty_{l=1}\alpha_lx^{2l}+
\sum\limits^\infty_{m=1}\beta_my^{2m}+
\sum\limits^\infty_{l=1}\sum\limits^\infty_{m=1}a_{lm}x^{2l}y^{2m}.
\eqno(4.15)
$$
Substituting (4.15) into (4.5'), we obtain the following equation:
$$
\begin{array}{rl}
\sum\limits^\infty_{l=1}\alpha_lx^{2l}+&
\sum\limits^\infty_{m=1}\beta_my^{2m}+
\sum\limits^\infty_{l=1}\sum\limits^\infty_{m=1}a_{lm}x^{2l}y^{2m}=\\
&C\epsilon\Delta-C\epsilon{(1+TC)}^{-1}x^2y^2 \\
+&C\epsilon{(1+TC)}^{-1}\left\{\sum\limits^\infty_{l=1}(\Delta x^{2l}-x^{2(l+1)}
y^2)\alpha_l\right. \\
+&\sum\limits^\infty_{m=1}(\Delta y^{2m}-x^2y^{2(m+1)})\beta_m \\
+&\sum\limits^\infty_{l=1}\sum\limits^\infty_{m=1}(\Delta x^{2l}y^{2m}-
\left.x^{2(l+1)}y^{2(m+1)})a_{lm}\right\}.
\end{array}
\eqno(4.16)
$$
For any $ x^{2l}y^{2m} $, after integration over $ t $, we have
$$
cx^{2l}y^{2m}=\f{1}{g}\int\limits^T_{-\infty}dt^\prime{(x_Te^{-T})}^{2l}
{(y_Te^{-bT})}^{2m}e^{(2l+2bm)t^\prime}=\f{1}{g}\f{1}{2(l+mb)}x^{2l}y^{2m}
\eqno(4.17)
$$
and
$$
(-TC)x^{2l}y^{2m}=\f{1}{2g}\f{1}{2(l+mb)}\left\{2l(2l-1)x^{2(l-1)}y^{2m}
+2m(2m-1)x^{2l}y^{2(m-1)}\right\}.
\eqno(4.18)
$$
Specially, we define, similarly to in ref.[2],
$$
\begin{array}{c}
C{(-TC)}^nx^{2l}=\Gamma^{l-n}_{l,(x)}x^{2(l-n)}~~~~(n<l) \\
C{(-TC)}^ny^{2m}=\Gamma^{m-n}_{m,(y)}y^{2(m-n)}~~~~(n<m) \\
\end{array}
\eqno(4.19)
$$
and
$$
\begin{array}{c}
C{(-TC)}^lx^{2l}=C\Gamma^1_{l,(x)} \\
C{(-TC)}^my^{2m}=C\Gamma^1_{m,(y)} \\
C{(-TC)}^{l+m}x^{2l}y^{2m}=C\Gamma^{1,1}_{l,m}.
\end{array}
\eqno(4.20)
$$
The expressions of the $ \Gamma $'s are given in Appendix.

Expanding
$$
{(1+TC)}^{-1}=1+(-TC)+{(-TC)}^2\cdots,
\eqno(4.21)
$$
applying (4.17) and (4.18) successively in (4.16), equating the
coefficients of each term $ x^{2l}y^{2m} $, a series of equations for
the unknown coefficients $ \{\alpha_l\}, \{\beta_m\} $ and $ \{a_{lm}\} $ are
obtained. To solve these equations we need to further expand $ \Delta $
and all the unknown coefficients on $ \epsilon $:
$$
\begin{array}{rl}
\epsilon\Delta=&\epsilon\Delta (1)+\epsilon^2\Delta(2)+\cdots \\
\alpha_l=&\epsilon\alpha_l (1)+\epsilon^2\alpha_l(2)+\cdots \\
\beta_m=&\epsilon\beta_m (1)+\epsilon^2\beta_m(2)+\cdots \\
a_{lm}=&\epsilon a_{lm} (1)+\epsilon^2a_{lm}(2)+\cdots \\
\end{array}
\eqno(4.22)
$$

First the coefficient of the $ x^0y^0 $-term gives
$$
\begin{array}{rl}
\Delta&-\Gamma^{1,1}_{1,1}+\sum\limits^\infty_{l=1}\sum\limits^\infty_{m=1}
(\Delta\Gamma^{1,1}_{l,m}-\Gamma^{1,~~~1}_{l+1,m+1})a_{lm} \\
&+\sum\limits^\infty_{l=1}(\Delta\Gamma^1_{l,(x)}-
       \Gamma^{1,~~1}_{l+1,1})\alpha_l \\
&+\sum\limits^\infty_{m=1}(\Delta\Gamma^1_{m,(y)}-
       \Gamma^{1,~1}_{1,m+1})\beta_m=0
\end{array}
\eqno(4.23)
$$

Introducing the $ \epsilon $-expansion, comparing the coefficient of the
same order of $ \epsilon $, we have, for $ \epsilon^1 $-term,
$$
\Delta(1)=\Gamma^{1,1}_{1,1}=\f{1}{4g^2b}
\eqno(4.24)
$$
$$
\begin{array}{rl}
\Delta(2)=&-\sum\limits^{\infty}_{l=1}\sum\limits^{\infty}_{m=1}
             (\Delta(1)\Gamma^{1,1}_{l,m}-
              \Gamma^{1,~~~1}_{l+1,m+1})a_{lm}(1) \\
          &-\sum\limits^{\infty}_{l=1}(\Delta(1)\Gamma^1_{l,(x)}-
              \Gamma^{1,~~1}_{l+1,1})\alpha_l(1) \\
          &-\sum\limits^{\infty}_{m=1}(\Delta(1)\Gamma^1_{m,(y)}-
              \Gamma^{1,~1}_{1,m+1})\beta_m(1) \\
          &\vdots \\
\end{array}
\eqno(4.25)
$$

To obtain $ \alpha_l(1), \beta_m(1) $ and $ a_{lm}(1) $ we analyze the
other terms in (4.16). After expansion on $ \epsilon $, the $ \epsilon^1 $
-term gives
$$
\begin{array}{c}
\sum\limits^\infty_l\alpha_l(1)x^{2l}+
\sum\limits^\infty_{m=1}\beta_m(1)y^{2m}+
\sum\limits^\infty_{l=1}\sum\limits^\infty_{m=1}a_{lm}(1)x^{2l}y^{2m} \\
=-Cx^2y^2-C(-TC)x^2y^2.
\end{array}
\eqno(4.26)
$$
Since
$$
\begin{array}{c}
-Cx^2y^2=-\f{1}{g}\f{1}{2(1+b)}x^2y^2 \\
-C(-TC)x^2y^2=-\f{1}{4g^2}\f{1}{1+b}(x^2+\f{y^2}{b})
\end{array}
\eqno(4.27)
$$
we have
$$
\begin{array}{rl}
\alpha_1(1)=&-\f{1}{4g^2}\f{1}{1+b},~~~~\alpha_{l>1}(1)=0 \\
\beta_1(1)=&-\f{1}{4g^2}\f{1}{b(1+b)}, ~~~~\beta_{m>1}(1)=0 \\
a_{11}(1)=&-\f{1}{g}\f{1}{2(1+b)},~~~~a_{lm}(1)=0~~ {\rm for}~~ l+m>2. \\
\end{array}
\eqno(4.28)
$$
Substituting (4.28) into (4.23), we have
$$
\Delta(2)=-\f{1}{16g^5b^3(1+b)}(b^2+4b+1)
\eqno(4.29)
$$

Now we turn to the $ \epsilon^2 $-term, keeping only the $ x^2 $- and
$ y^2 $-terms we have
$$
\begin{array}{rl}
\alpha_1(2)x^2+\beta_1(2)y^2=&\Delta(1)C(\alpha_1(1)x^2+\beta_1(1)y^2) \\
&-C{(-TC)}^2(\alpha_1(1)x^4y^2+\beta_1(1)x^2y^4) \\
&+\Delta(1)C(-TC)a_{11}(1)x^2y^2 \\
&-C{(-TC)}^3a_{11}(1)x^4y^4
\end{array}
\eqno(4.30)
$$
Applying (4.17) and (4.18) successively, substituting (4.24), (4.28)
into (4.30), comparing the coefficients of $ x^2 $-term and $ y^2
$-term, respectively, we have
$$
\begin{array}{rl}
\alpha_1(2)=&\f{1}{16g^5{(1+b)}^2}[\f{b+2}{b}+\f{18}{(2+b)(1+2b)}
+\f{3}{2+b}] \\
\beta_1(2)=&\f{1}{16g^5{(1+b)}^2}[\f{2b+1}{b^3}+\f{18}{b(2+b)(1+2b)}+
\f{3}{b^2(1+2b)}].
\end{array}
\eqno(4.31)
$$
Following similar procedure for the other terms of the order $ \epsilon^2 $,
we obtain the coefficients

\hs*{5mm}\bm{100mm}
$ a_{11}(2)=\f{1}{8g^4{(1+b)}^2}[\f{5}{2b}+\f{18}{(1+2b)(2+b)}] $,~~~~~ for
$ x^2y^2 $ term;

$ \alpha_2(2)=\f{4+b}{32g^4{(1+b)}^2(2+b)} $, ~~~~~~~~~~for $ x^4 $-term;

$ \beta_2(2)=\f{1+4b}{32g^4{(1+b)}^2(1+2b)b^2} $, ~~~~~~~~for $ y^4 $-term;

$ a_{21}(2)=\f{4+b}{8g^3{(1+b)}^2(2+b)} $, ~~~~~~~~~~for $ x^4y^2 $-term;

$ a_{12}(2)=\f{1+4b}{8g^3b{(1+b)}^2(1+2b)} $, ~~~~~~~~~~for $ x^2y^4
$-term.\vs*{4mm}\em

\noindent
and

\hs*{5mm} $ a_{22}(2)=\f{1}{8g^2{(1+b)}^2} $, ~~~~~~~~~~~for $ x^4y^4 $-term.

\noindent
Comparing to (3.19) the above results are exactly the same up to the order
of $ \epsilon^2 $ and $ g^{-5} $.

\section*{\bf 5. Summary and conclusion}
\setcounter{section}{5}
\setcounter{equation}{0}

The single trajectory quadrature method newly developed in Refs.[1,2] is applied
to solve the ground state wave function and energy for Schroedinger
equation with a two-dimensional non-separable potential.

Different versions of perturbation expansion, as well as the Green's
function based on this new method are tested using this example. The
results are also compared to the one from the traditional perturbation
theory (see Appendix B). The consistency of all the results proves the
applicability and potential of the various versions of this new method.
To solve a problem based on the new
perturbation expansion method it seems better to use $ g\lambda U $
as the perturbed potential. Compared to the potential $ \epsilon U $,
a faster convergence could be obtained; while compared to the potential
$ g^2\mu U$, $g\lambda U $ could avoid the complex in calculating
the trajectory $ {\bf S}_0 $ when $ \mu U $ is included in $ v $, as
shown in section II. While the perturbation expansion seems to rely on
the scale factor $ g $, the Green's function method is more general and
provides a much wider applicability. It is of interests to apply this new
method to other areas of physics where Schroedinger
equation with strong potential or with various kinds of perturbations
needs to be solved.

The authors would like to thank Professor T. D. Lee for his continuous
instructions and advice. This work is partly supported by NNSFC (No. 19947001).

\bc
{\bf Reference}
\ec

1. R. Friedberg, T. D. Lee and W. Q. Zhao, IL Nouvo Cimento A112, 1195(1999)

2. R. Friedberg, T. D. Lee and W. Q. Zhao, Ann. Phys. 288, 52(2001)

3. W. C. Lee and T. K. Lee, Preprint

4. H. Zhai, J. F. Liao, P. F. Zhuang and W. Q. Zhao, Preprint

\hspace*{.3cm} P. P. Yu and H. Guo, Preprint

\vs*{.5cm}

\bc
{\bf Appendix A}
\ec
In the calculation of (4.16) the following expressions are needed
$$
\begin{array}{rl}
Cx^{2l}=&\f{1}{g}\f{1}{2l}x^{2l} \\
(-TC)x^{2l}=&\f{1}{2g}(2l-1)x^{2(l-1)} \\
Cy^{2m}=&\f{1}{g}\f{1}{2mb}y^{2m} \\
(-TC)y^{2m}=&\f{1}{2g}{\f{1}{b}}(2m-1)y^{2(m-1)} \\
\Gamma^1_{l,(x)}=&\f{(2l-1)!!}{{(2g)}^l} \\
\Gamma^1_{m,(y)}=&\f{(2m-1)!!}{{(2gb)}^m}
\end{array}
$$

Expression of some $ \Gamma^{1,1}_{l,m} $ applied in our paper are given
in the following:
$$
\begin{array}{rl}
\Gamma^{1,1}_{1,1}=&\f{1}{4g^2b} \\
\Gamma^{1,1}_{2,1}=&\f{1}{8g^3(2+b)}(\f{6}{b}+3) \\
\Gamma^{1,1}_{1,2}=&\f{1}{8g^3(1+2b)}(\f{3}{b^2}+\f{6}{b}) \\
\Gamma^{1,1}_{2,2}=&\f{1}{32g^4(1+b)}[\f{6}{1+2b}(\f{3}{b^2}+\f{6}{b})+
\f{6}{2+b}(\f{6}{b}+3)] \\
\end{array}
$$

\newpage

\bc
{\bf Appendix B}
\ec
For comparison the result for the same Hamiltonian based on the traditional
perturbation  theory is given in the following.

For the Hamiltonian
\begin{eqnarray}
H&=&H_0 +\epsilon U\nonumber\\
H_0&=& -\f{1}{2}\nabla^2 + \f{1}{2} g^2(x^2+b^2y^2)\nonumber\\
\epsilon U&=& x^2+y^2\nonumber
\end{eqnarray}
we introduce
\begin{eqnarray}
u_m^{(\omega)}(s)=(2^m~m!\sqrt{\f{\pi}{\omega}})^{-\f{1}{2}}~
e^{-\f{1}{2}\omega x^2}~H_m(\sqrt{\omega}~s),\nonumber
\end{eqnarray}
where
\begin{eqnarray}
H_m(\xi)=(-1)^m~e^{-\xi^2}\f{d^m~e^{-\xi^2}}{d\xi^m}\nonumber
\end{eqnarray}
and
\begin{eqnarray}
\omega=&g~~~&{\rm for}~~s=x\nonumber\\
       &gb~~~&{\rm for}~~s=y.\nonumber
\end{eqnarray}
The unperturbed part gives
\begin{eqnarray}
H_0~\psi_{m~n}^{(0)}(x,y)=E_{m~n}^{(0)}~\psi_{m~n}^{(0)}(x,y),\nonumber
\end{eqnarray}
where the unperturbed wave function and energy are
\begin{eqnarray}
\psi_{m~n}^{(0)}(x,y)&=&u_m^{(g)}(x)~u_n^{(gb)}(y)\nonumber\\
E_{m~n}^{(0)}&=&\f{1}{2}(gm+gbn).\nonumber
\end{eqnarray}
Introducing matrix elements
\begin{eqnarray}
F_{m,~n}^{(\omega)}=\int ds~u_m^{(\omega)}(s)~s^2~u_n^{(\omega)}(s),
\nonumber
\end{eqnarray}
the values of $F_{m,~n}^{(\omega)}$, relevant to our problem, are given in
the following table.

\vspace{.5cm}

\begin{tabular}{rlccc}
   &m~~&~~~~~0~~~~~&~~~~~2~~~~~&~~~~~4~~~~~\\
n&&&&\\
&&&&\\
0&&$\frac{1}{2\omega}$&$\frac{1}{\sqrt{2}\omega}$&~~~~~0~~~~~\\
&    &&&\\
2&&$\f{1}{\sqrt{2}\omega}$&$\f{5}{2\omega}$&$\f{\sqrt{3}}{\omega}$\\
    &&&&\\
4&&0&$\f{\sqrt{3}}{\omega}$&\\
\end{tabular}

\vspace{.5cm}

Up to the order  of $\epsilon^2$, the perturbed energy and ground state
wave function could be expressed as
\begin{eqnarray}
&\epsilon \Delta E_{0~0}^{(1)}+\epsilon^2 \Delta E_{0~0}^{(2)}
\nonumber\\
&\epsilon \psi_{0~0}^{(1)}(x,~y)+\epsilon^2 \psi_{0~0}^{(2)}(x,~y).
\nonumber
\end{eqnarray}
Based on the basic formula of  the traditional perturbation theory, using
the values of $F_{m,~n}^{(\omega)}$ given  in the table, we have
\begin{eqnarray}
\Delta E_{0~0}^{(1)}&=&<\psi_{0~0}^{(0)}|x^2y^2|\psi_{0~0}^{(0)}>\nonumber\\
                    &=&F_{0~0}^{(g)}~F_{0~0}^{(gb)}=\f{1}{4g^2b},\nonumber
\end{eqnarray}
\begin{eqnarray}
\Delta E_{0~0}^{(2)}&=&\sum\limits_{m~n}~'~\f
{|<\psi_{0~0}^{(0)}|x^2y^2|\psi_{m~n}^{(0)}>|^2}
{E_{0~0}^{(0)}-E_{m~n}^{(0)}}
          \nonumber\\
                    &=&-\sum\limits_{m~n}~'~
                    \f{(F_{0~m}^{(g)}~F_{0~n}^{(gb)})^2}{mg+ngb}
          \nonumber\\
                    &=&-\f{1}{16g^5b^3(b+1)}(b^2+4b+1),\nonumber
\end{eqnarray}
\begin{eqnarray}
\psi_{0~0}^{(1)}(x,y)&=&\sum\limits_{m~n}~'~\f
{<\psi_{0~0}^{(0)}|x^2y^2|\psi_{m~n}^{(0)}>}
{E_{0~0}^{(0)}-E_{m~n}^{(0)}}\psi_{m~n}^{(0)}(x,y)
          \nonumber\\
                    &=&-\sum\limits_{m~n}~'~
                    \f{F_{0~m}^{(g)}~F_{0~n}^{(gb)}}{mg+ngb}
                    \psi_{m~n}^{(0)}(x,y)
          \nonumber\\
          &=&\psi_{0~0}^{(0)}(x,y)\{\f{(b^2+b+1)}{8g^3b^2(b+1)}
             -\f{1}{4g^2(b+1)}(x^2+\f{y^2}{b})-\f{x^2y^2}{2g(b+1)}\}
          \nonumber
\end{eqnarray}
and
\begin{eqnarray}
\psi_{0~0}^{(2)}(x,y)&=&\sum\limits_{m'~n'}~'~\{\sum\limits_{m~n}~'~
\f{<\psi_{0~0}^{(0)}|x^2y^2|\psi_{m~n}^{(0)}>
<\psi_{m~n}^{(0)}|x^2y^2|\psi_{m'~n'}^{(0)}>}
{(E_{0~0}^{(0)}-E_{m~n}^{(0)})(E_{0~0}^{(0)}-E_{m'~n'}^{(0)})}\nonumber\\
&&-\f{<\psi_{0~0}^{(0)}|x^2y^2|\psi_{0~0}^{(0)}>
<\psi_{0~0}^{(0)}|x^2y^2|\psi_{m'~n'}^{(0)}>}
{(E_{0~0}^{(0)}-E_{m'~n'}^{(0)})^2} \}
\psi_{m'~n'}^{(0)}(x,y)
          \nonumber\\
&=&\sum\limits_{m'~n'}~'~\{\sum\limits_{m~n}~'~
   \f{F_{0~m}^{(g)}~F_{0~n}^{(gb)}F_{m~m'}^{(g)}~F_{n~n'}^{(gb)}}
   {(mg+ngb)(m'g+n'gb)}
-  \f{F_{0~0}^{(g)}~F_{0~0}^{(gb)}F_{0~m'}^{(g)}~F_{0~n'}^{(gb)}}
   {(m'g+n'gb)^2} \}
\psi_{m'~n'}^{(0)}(x,y)
          \nonumber\\
&=&\f{1}{g^6} \{\nonumber\\
&&\f{1}{16\sqrt{2}~b^3(b+1)}(2b^2+8b+1)\psi_{2~0}^{(0)}(x,y)
          \nonumber\\
&& \f{1}{16\sqrt{2}~b^4(b+1)}(b^2+8b+2)\psi_{0~2}^{(0)}(x,y)
          \nonumber\\
&& +\f{1}{32~b^3(b+1)^2}(5b^2+34b+5)\psi_{2~2}^{(0)}(x,y)
          \nonumber\\
&& +\f{\sqrt{3}}{16~b^2(b+1)(b+2)}(b+6)\psi_{4~2}^{(0)}(x,y)
          \nonumber\\
&& +\f{\sqrt{3}}{16~b^3(b+1)(2b+1)}(6b+1)\psi_{2~4}^{(0)}(x,y)
          \nonumber\\
&& +\f{\sqrt{3}}{32\sqrt{2}~b^2} \cdot \f{b+3}{b+1}\psi_{4~0}^{(0)}(x,y)
          \nonumber\\
&& +\f{\sqrt{3}}{32\sqrt{2}~b^4} \cdot \f{3b+1}{b+1}\psi_{0~4}^{(0)}(x,y)
          \nonumber\\
&& +\f{3}{16~b^2(b+1)^2}\psi_{4~4}^{(0)}(x,y)\}.
          \nonumber
\end{eqnarray}
Substituting the expression of $\psi_{0~0}^{(0)}(x,y)$,
keeping the terms up to the order of $g^{-5}$, writing  the obtained
wave function as
\begin{eqnarray}
&&\psi_{0~0}^{(0)}(x,y)(1+\f{b^2+b+1}{8g^3b^2(b+1)})\cdot\chi(x,~y)\nonumber\\
&\propto& e^{-\f{1}{2}g(x^2+by^2)}~\chi(x,~y)\nonumber\\
&=&e^{-g~S_0}\cdot \chi(x,~y).\nonumber
\end{eqnarray}
Comparing the coefficients of each $x^my^n$ term, the obtained
$\chi(x,~y)$ is exactly the same as the results based on
the newly developed method shown in our paper.

\end{document}